\documentclass{emulateapj}

\usepackage{amsmath, amsthm, amssymb, graphics}
\usepackage{graphicx}
\usepackage{natbib}
\usepackage{color}

\shorttitle{Tomographic reconstruction of the three-dimensional structure of the HH30 jet}
\shortauthors{F. De Colle et al.}

\begin{document}

\title{Tomographic reconstruction of the three-dimensional structure of the HH30 jet}

\author{F. De Colle\altaffilmark{1}}
\affil{Astronomy \& Astrophysics Department, University
of California, Santa Cruz, CA 95064, USA}
\email{fabio@ucolick.org}

\author{C. del Burgo\altaffilmark{2}}
\affil{UNINOVA-CA3, Campus da Caparica, Quinta da Torre,
Monte de Caparica 2825-149, Caparica, Portugal}
\email{cburgo@uninova.pt}

\and

\author{A. C. Raga\altaffilmark{3}}
\affil{Instituto de Ciencias Nucleares, Universidad Nacional Aut\'onoma de M\'exico, 
Ap.P. 70543, 04510 DF, Mexico}
\email{raga@nucleares.unam.mx}

\date{Received xxx; accepted xxx}

\begin{abstract}
  The physical parameters of Herbig-Haro jets are usually 
  determined from emission line ratios, obtained from spectroscopy or narrow band
  imaging, assuming that the emitting region is homogeneous along the line of sight.
  Under the more general hypothesis of axisymmetry, we apply
  tomographic reconstruction techniques to the analysis of Herbig-Haro jets. 
  We use data of the HH30 jet taken by \citet{2007ApJ...660..426H}
  with the Hubble space telescope using the slitless spectroscopy technique.
  Using a non-parametric Tikhonov regularization technique,
  we determine the volumetric emission line intensities of the
  [\ion{S}{2}]$\lambda\lambda$6716,6731, [\ion{O}{1}]$\lambda$6300 
  and [\ion{N}{2}]$\lambda$6583 forbidden emission lines.
  From our tomographic analysis of the corresponding line ratios, we produce 
  ``three-dimensional'' images of the physical parameters.
  The reconstructed density, temperature and ionization fraction present much
  steeper profiles than those inferred using the assumption of 
  homogeneity. Our technique reveals that the reconstructed jet is much more collimated 
  than the observed one close to the source (a width $\sim 5$~AU vs. 
  $\sim 20$~AU at a  distance of 10~AU from the star), while they have 
  similar widths at larger 
  distances. In addition, our results show a much more fragmented
  and irregular jet structure than the classical analysis, 
  suggesting that the the ejection history of the jet from 
  the star-disk system has a shorter timescale component ($\sim$ some months) 
  superimposed on a longer, previously observed timescale (of a few years).
  Finally, we discuss the possible application of the same technique to 
  other stellar jets and planetary nebulae.
\end{abstract}




\section{Introduction}

Herbig-Haro (HH) jets appear as narrow and well 
collimated ($\sim$ a few degrees) knotty beams of atomic/ionic
and/or molecular gas moving away from the source with supersonic 
velocities ($\sim 100-300$~km s$^{-1}$) (e.g. \citealt{2001ARA&A..39..403R}).

While considerable progress has been obtained in the understanding
of the behavior of HH jets, long standing problems 
remain unsolved, including the 
origin of the stellar jets themselves (e.g. \citealt{1997A&A...319..340F,2000prpl.conf..789S}), 
the importance of jets in solving the angular momentum 
problem (e.g. \citealt{2009pjc..book...23H}), 
the origin of knots along stellar jets (e.g. \citealt{2008ApJ...688.1137D}), 
the feedback of jets on the core and the star formation efficiency, 
and the interaction of jets
with the ambient medium eventually leading to the generation of 
turbulence in molecular clouds (e.g. \citealt{2009ApJ...695.1376C}).
To properly understand these phenomena, it is crucial to 
accurately determine the physical conditions of the jet plasma.

The proper determination of the electron density $n_\mathrm{e}$, 
temperature $T_\mathrm{e}$, and hydrogen ionization
fraction $x_\mathrm{H}$ gives strong insights, for instance, on 
the nature of the heating source of the jet material, on the mechanism 
that leads to the production of the knots and on the interaction 
of the outflowing material with the surrounding environment.
The determination of the total density $n_\mathrm{H}$ 
($=n_\mathrm{e}/x_\mathrm{H}$) is important in order to
derive the mass flux, and more significantly, to understand the role
played by jets in the removal of angular momentum from the
star and the circumstellar disk. Furthermore, different jet ejection models
predict different dependencies of $n_\mathrm{H}$ as a function 
of the distance from the jet axis 
(e.g. \citealt{1995ApJ...455L.155S,2000ApJ...533..897L}).

Two approaches are used to infer the outflow excitation conditions 
from the observations.
The physical parameters can be calculated directly from the observed
line ratios (e.g. \citealt{1999A&A...342..717B};
\citealt{2005A&A...441..159N}; \citealt{2006A&A...456..189P}; 
\citealt{2007ApJ...660..426H}, hereafter HM07).
Alternatively, the observations can be compared with 
predictions of emission line intensities from models of different complexity,
ranging from ``simple'' parallel shock models (e.~g. \citealt{1987ApJ...316..323H}, 
\citealt{1994ApJ...436..125H}, \citealt{2000A&A...356L..41L}, 
\citealt{2003A&A...410..155P}) to complex multi-species numerical simulations 
(e.g. \citealt{2007A&A...465..879R}).
All these studies agree that the jet is denser close to the source, 
with temperatures and ionization fractions which, although peaking 
at the positions of the knots, tend to decrease for larger distances
from the source.

While previous studies have focused on the variation of the
physical conditions along the jet axis,
recent imaging of resolved jets has been performed
(e.g. \citealt{2000ApJ...537L..49B, 2002ApJ...576..222B, 
2004ApJ...604..758C, 2007ApJ...663..350C, 
2005A&A...432..149W, 2007AJ....133.1221B}, HM07).
It has been observed in these studies that the electron and total 
densities peak on the axis of the jet and decrease away 
from the axis (within e.g. a factor of 2 in HM07). The temperature
and ionization fraction cross-sections are found to be more homogeneous.

We note, however, that the above-mentioned findings were obtained
assuming that stellar jets are homogeneous along the line of sight.
In a previous paper \citep[hereafter Paper I]{2008A&A...485..765D}
we discussed in detail the error in the determination of the
physical parameters introduced when assuming a jet as homogeneous
along the line of sight.
In Paper I we proposed the application of tomographic techniques to
determine the volumetric physical parameters, showing an example of the application
of a parametric ``multi-Gaussian'' \citep{1991ApJ...366..599B} method to reconstruct
the cross-section of the HH30 jet at one particular position along the jet.
In this paper, we extend the previous results by reconstructing the full two 
dimensional structure of the HH30 jet by implementing a non-parametric Tikhonov 
reconstruction technique.

In particular, Section 2 explains in detail the problems related
to the inversion of the physical parameters, and
the methods employed in the inversion problem. Section 3
presents the application of the reconstruction process to the HH30 jet.
Section 4 and 5 present a discussion of the results and our conclusions.


\section{The determination of the jet structure}

\subsection{General considerations}

We consider a jet with known physical parameters
(hereafter, we refer to the jet electron density $n_\mathrm{e}$, 
electron temperature $T_\mathrm{e}$ and ionization 
fraction $x_\mathrm{H}$ as the physical parameters of the jet).
To determine synthetic emissivities from the physical parameters 
it is necessary to follow a series of steps 
(e.g. \citealt{2010LNP...793..213D}):
\begin{enumerate}
\item The emission coefficients are calculated, assuming
      statistical equilibrium and determining the population of
      each level for each considered atom/ion.
\item The emission coefficients are integrated along the line of sight.
\item The synthetic emissivity are convolved with the seeing (i.e., 
      blurring and scintillation of the jet in the Earth's atmosphere), 
      the response of the detector, the sampling design and the instrumental
      point spread function (PSF).
\end{enumerate}

The inverse process consists of deriving the  
emission coefficients from the observed emission 
lines, and determining the physical parameter from the ratio of
the emission coefficients.
Actually, the observed emission lines always have an associated
intrinsic noise.
As a consequence, deconvolution and inversion processes are in
general ``ill-conditioned'' problems (i.e. similar initial conditions
lead to very different deconvolved/inverted solutions).
The level of ill-conditioning depends on
the particular form of the convolution/integration operator (``the kernel''), 
and on the level of noise in the data (e.g. \citealt{1986ipag.book.....C, 
1995InvPr..11..783B}).

The convolution also produces a loss of
information, that again depends on the form of the kernel.
In fact, the convolution process may be seen as a low-pass filter.
All the structures with size smaller than the characteristic
size of the kernel (the PSF/seeing) will not be reconstructed
properly when deconvolving the image.

In addition, the determination of the physical parameters
from the emission coefficients depends on several assumptions
and uncertainties as, for instance, those associated with the ratios of the
populations of the emitting species  \citep{2006A&A...456..189P}.

Finally, to solve the tomography problem, it is necessary to make 
some assumptions about the geometry of the problem. 
In some cases, in stellar jets, a reasonable hypothesis
is to consider the jet as cylindrically symmetric.
This hypothesis describes better the three dimensional structure 
of the jet than the commonly used assumption of homogeneity
along the line of sight.
 
\subsection{Reconstruction of the Jet Structure Using Tomographic Techniques}
\label{sec:met}

Let us consider an axisymmetric jet moving in the plane of the sky.
In the following, we use a two-dimensional Cartesian reference system
centered on the star-disk system, and defined on the plane of the sky, 
such that the $x$- and $y$-axes are perpendicular and parallel
(respectively) to the ``main'' axis of the jet.
Additionally, we indicate with $r$ the cylindrical
distance from the jet axis.

The aim of the inversion process is to 
determine the emission coefficients $i(r,y)$ as a function
of the observed intensities $I(x,y)$.
The problem is further reduced to one dimension 
assuming that the variation of the physical parameters and 
therefore of the emission lines along the $y$-axis (parallel
to the direction of propagation of the jets) is much smaller 
than the variation in the direction across the jet 
(i.e., along the $r$, $x$-axis). 

With these approximations, the observed emission line intensity cross section
$I(x)$ is related to the emission coefficient $i(r)$ by the Abel transform:
\begin{equation}
  I(x) = 2\int_x^{R} \frac{i(r)r dr}{\sqrt{r^2-x^2}},
  \label{eq:abel}
\end {equation}
where $R$ is the jet radius.
equation (\ref{eq:abel}) represents an approximation of the relation between
observed and volumetric emission coefficients (see Appendix A).

As mentioned previously, the inversion of equation (\ref{eq:abel}) constitutes
an ill-posed problem. 
The reason for that is easily
seen looking at the analytical solution of the Abel transform, given by:
\begin{equation}
  i(r) = -\frac{1}{\pi} \int_r^{R} \frac{d I}{dx}
             \frac{dx}{\sqrt{x^2-r^2}} . 
 \label{eq:ilp}
\end {equation}
Decomposing the data into a continuous noise-free
and an oscillatory noisy part (expanded as a function of the
frequencies $\omega_n$) $I = I_0 + \delta \sum_n \sin \omega_n x$,
the derivative is:
$dI/dx = dI_0/dx + \delta \sum_n \omega_n \sin \omega_n x$, and the high 
frequency noise dominates the integral of equation (\ref{eq:ilp}).

Equation (\ref{eq:abel}) can be written as
\begin{equation}
d = A x\;,
\label{eq:axd}
\end{equation}
where $x$ is a vector of $M$ elements (a discretization of the solution 
$i(r)$ of the inverse problem), 
$d$ is a vector of $N$ elements representing the data $I(x_i)$, and $A$ is
a matrix with dimension $M \times N$. $A$ is an operator that 
transforms the volumetric into the integrated intensities, and 
is given by a particular discretization of the integral present in 
equation (\ref{eq:abel}) (see Appendix A).

The least-squares solution of this linear system of equations 
is obtained minimizing the norm of the residuals:
\begin{equation}
\min \|A x- d\|_2^2 \;,
\label{eq:nor}
\end{equation}
where the standard definition of the 2-norm of a vector of $n$ elements
is given by 
\begin{equation}
\|x\|_2 = \sqrt{x_1^2 + x_2^2 + \dots + x_n^2}\;.
\end{equation}

For noisy data, the solution determined minimizing the norm of the 
residuals (eq. [\ref{eq:nor}]), as shown before, is dominated by high frequency noise.
A standard solution to this problem (e.g., \citealt{1986ipag.book.....C, tombook}) is obtained 
by ``regularizing'' the data, which in the Tikhonov approach consists
of solving the following damped least squares problem:
\begin{equation}
\min \left(\|A x- d\|_2^2 + \alpha^2 \| L x \|_2^2 \right) \;,
\label{eq:tik}
\end{equation}
with the condition:
\begin{equation}
\|A x- d\| \le \delta \;,
\label{eq:alp}
\end{equation}
where $\alpha$ is the ``regularization parameter'', $L$ is the 
``Tikhonov matrix'', and $\delta$ is the uncertainty (i.e., the noise level) 
in the data.
$L$ can be chosen to minimize the norm of the solution, or
the first or second derivatives of the solution (see Appendix A).

Basically, the Tikhonov method chooses the smoothest solution 
(with respect to the zero-, first-, second-, etc. order derivative, 
depending on the particular form of $L$ used)
such that the residuals are of the order of the noise in the data.
The value of $\alpha$ obtained by equation (\ref{eq:alp}) determines
the ``optimal'' (i.e. the smoothest) solution compatible with
the data and their associated errors.
In particular, while a too small value of $\alpha$ produces a solution dominated 
by noise, a larger value of $\alpha$ produces a too smooth 
solution.
Finally, it is important to note that the solution of the inverse
problem depends on the particular form of the Tikhonov
matrix $L$ (that is a free parameter of the problem). This is related 
to the well-known fact that, to solve an inverse problem, it is
necessary to introduce a certain amount of information on the 
form of the solution (e.g., on the smoothness of the solution 
or its derivative of order $n$).

Once the emission coefficients have been reconstructed, the 
physical parameters are determined from their ratios.


\section{Results}


\subsection{The HH30 Data}

The HH30 jet is one of the best studied outflows from T Tauri stars
(\citealt{1991A&A...252..740M, 1996ApJ...473..437B, 1996ApJ...468L.103R, 
1999A&A...350..917B, 2006A&A...458..841P, 2007AJ....133.2799A}, HM07).
It is located at a distance of 140 pc \citep{1994AJ....108.1872K}, in the Taurus molecular cloud.
HH30 is nearly on the plane of the sky, within an angle of
$\sim 1^\circ$, and it is approximately axisymmetric.
Because the outflow lies so close to the plane of the sky,
the flared disk absorbs a large part of the radiation
coming from the star, allowing the study of the jet at distances
$\gtrsim 20$~AU from the star.
The jet is bipolar, but with a blueshifted lobe much more luminous
than the redshifted lobe.

Furthermore, high resolution, multi-epoch images of this jet
are available. The data, presented by HM07, were obtained by
the Space Telescope Imaging Spectrograph (STIS) on 
HST using a slitless spectroscopy technique, during 
two epochs, in 2000 and 2002. They consist of two-dimensional 
images of the jet in a series of emission lines, 
including four bright optical lines:
the [\ion{S}{2}]$\lambda\lambda$6716,6731 doublet, the  
[\ion{O}{1}]$\lambda$6300 line and the [\ion{N}{2}]$\lambda$6583 line.

In the images, the cross section of the jet (i.e., 
in the direction perpendicular to the jet axis) is resolved 
with $\sim 15-20$ pixels, depending on the distance from 
the central star.  
As determined by HM07, the PSF FWHM is of $\sim 4$ pixels, 
with each pixel size corresponding to 0.025\arcsec or 
3.55 AU/pixel.
As stated by HM07, the S/N is very high along the 
jet knots, with uncertainties below $5\%$.

\subsection{Line reconstruction}
\label{sec:lin}

The Tikhonov regularization technique
has been applied to the inversion of the observed data, 
to reconstruct the emission coefficients.

\begin{figure}
\centering
 \includegraphics[width=0.5\textwidth]{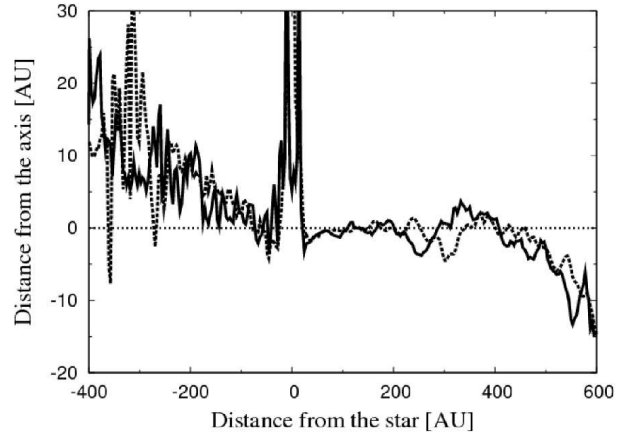}
\caption{Position of the jet axis $x_0$ relative 
to the star. The $x$ and $y$-axis 
are parallel and perpendicular to the main axis
of the jet, respectively. Negative/ positive $x$ 
distances correspond to the redshifted/blueshifted outflow.}
\label{fig:x0}
\vskip 0.5cm
\end{figure}

To invert equation (\ref{eq:abel}) it is necessary to know
the position $x_0$ of the jet axis in the emission line images.
As the HH30 jet is not perfectly axisymmetric,
$x_0$ is a slowly varying function of $y$, i.e. $x_0=x_0(y)$.
We determine $x_0(y)$ by a Gaussian fit to the total line 
intensity cross-section.
The resulting $x_0$ is shown in Fig. \ref{fig:x0}. 

The blueshifted jet bends 
at distances $\gtrsim 400$~AU, probably due 
to precession \citep{1996ApJ...473..437B}, while 
the redshifted jet (much more noisy due to
the lower S/N of the emission line images) it is
not aligned with its blueshifted counterpart, forming an 
angle of $\approx 2.5^\circ$.
It is unclear whether this asymmetry is due to 
inhomogeneities of the ambient medium or is a property of the
ejection mechanism, although the fact that it starts close
to the source seems to favor the latter explanation.

An asymmetry of the ejection mechanism itself could be 
related to an asymmetry in the stellar magnetic field.
However, exploratory three-dimensional simulations by 
\citet{2009MNRAS.399.1802R} of a stellar 
dipole magnetic field misaligned with the rotation axis of 
the star and the disk produce winds that are axisymmetric 
about the rotational axis.
More complex magnetic field configurations have been studied 
by \citet{2010arXiv1004.0385L}. These authors used two-dimensional
axisymmetric simulations to study the case of a 
rotating star with a large quadrupole component, showing 
that intrinsically asymmetric jets (with respect to the 
equatorial plane) can be generated in this case.
Therefore, a possible origin for the asymmetry could be a 
large quadrupole component of the stellar magnetic field,
misaligned with the star + disk rotation axis.

\begin{figure}
\centering
      \includegraphics[height=10cm]{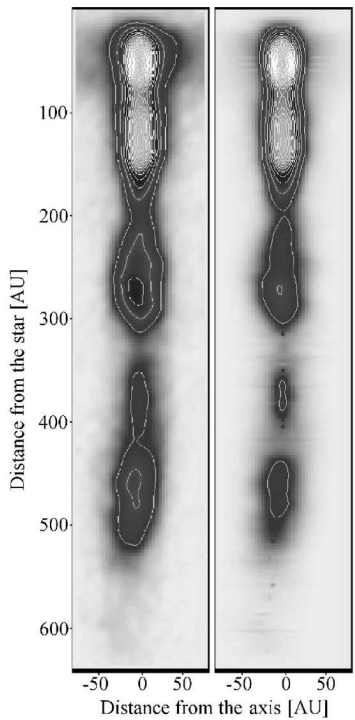}
\caption{Observed (left) and reconstructed (right) 
emission line images (arbitrary units).
The images refers to the sum of all emission lines
(reconstructed independently).
The central source is on the top of the Figure.}
\label{fig:lines}
\end{figure}

Consistently with the axisymmetry hypothesis, 
pixels symmetric with respect to $x_0$ are combined to produce 
symmetric emission-line cross-sections.
The emission coefficients are then determined inverting the Abel transform 
(eq. [\ref{eq:abel}]) for each emission line.
The resulting emission line images (Fig. \ref{fig:lines}) are sharper than 
the original, while still conserving the main features of the jet (e.g. the 
position and number of knots).

However, sharper emission lines do not necessarily imply 
different line ratios and therefore different physical parameters.
In fact (see Paper I), in the case of an observed emission
line with a gaussian dependence on $x$: $I(x) \propto \exp(-(x/\sigma)^2)$, 
the emission coefficients scale 
as $i(r) \propto 1/\sigma \exp(-(r/\sigma)^2)$.
The relation between the reconstructed and original line ratios is therefore
\begin{equation}
  \frac{i_1}{i_2} \propto \frac{\sigma_2}{\sigma_1} \frac{I_1}{I_2} \;.
  \label{eq:simpl}
\end{equation}

\begin{figure}
\centering
 \includegraphics[height=12cm]{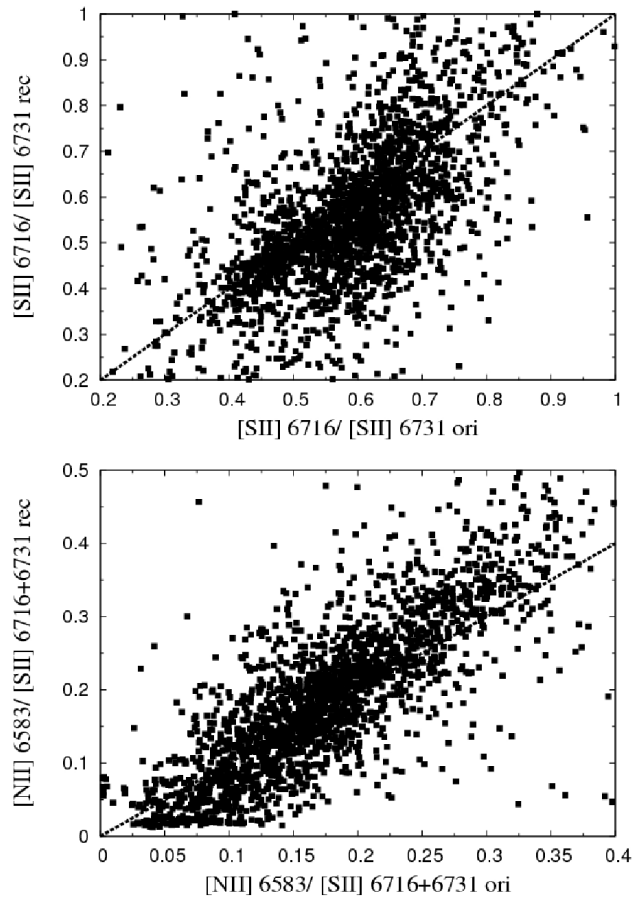}
\caption{Reconstructed vs. original 
[\ion{N}{2}]/[\ion{S}{2}] ratio (upper panel) and [\ion{S}{2}] 6716/[\ion{S}{2}] 6731
 ratio (lower panel). Only pixels with emission lines with S/N $>$ 10
are included. The dashed line shows the curve $y=x$.}
\label{fig:recratios}
\vskip 0.5cm
\end{figure}

The [\ion{S}{2}] 6716/[\ion{S}{2}] 6731 and [\ion{N}{2}] 6583/[\ion{S}{2}] 6716+6731 
reconstructed ratios are shown in fig. \ref{fig:recratios} as a function of
the observed ratios.

The [\ion{S}{2}] 6716/[\ion{S}{2}] 6731 reconstructed ratio is systematically
lower than the original one, with a larger effect for lower 
values of the ratio. 
As this ratio is a monotonic decreasing function of $n_\mathrm{e}$,
this result implies that the inversion process produces in general larger
electron density, with a larger increase for densities close to the
high density limit. 
The large spread in the data is not related to errors
in the tomographic inversion process. The inversion of a pixel with a 
certain [\ion{S}{2}] ratio is not unique, as it depends on the emission line 
cross-sections and not only on its own value.

The [\ion{N}{2}] 6583/[\ion{S}{2}] 6716+6731 ratio presents
a more complex behavior.
It depends on all the physical parameters (with a stronger dependence on 
$x_\mathrm{H}$), and it is a good shock tracer (HM07), as it becomes
higher for denser, hotter and more ionized gas.
Fig. \ref{fig:recratios} show that this reconstructed ratio 
increases (with respect to the original) for high values, while slightly
decreasing for low values, indicating that shocks will be more evident
in the reconstructed jet.

\subsection{The Determination of the Physical Parameters}
\label{sec:par}

To determine $n_\mathrm{e}$, $T_\mathrm{e}$ and $x_\mathrm{H}$ 
we use the BE method \citep{1999A&A...342..717B}. 
In this method, Nitrogen and Oxygen ionization 
fractions are determined assuming charge exchange 
equilibrium with Hydrogen, while Sulfur is assumed to 
be completely ionized. From four emission lines, three
independent ratios are formed. Being the ratios
just function of $n_\mathrm{e}$, $T_\mathrm{e}$ and 
$x_\mathrm{H}$ (fixing the relative population of S, O, N), 
these parameters can be determined easily 
by any root-finding algorithm.

This procedure has been improved recently by HM07,
including all the available ratios. 
HM07 showed that the best estimation of the physical 
parameters is the one obtained by the minimization 
of the least square problem 
\begin{equation}
\min \sum_{k=1}^p \frac{\left(r_k-m_k\right)^2}{\sigma_k^2}
\label{eq:bestfit}
\end{equation}
where $m_k=m_k(n_\mathrm{e},T_\mathrm{e},x_\mathrm{H})$ is the model prediction, 
$r_k$ is the natural logarithm of the observed $k-$ratio,
$p$ is the total number of ratios considered,
and $\sigma_k$ is the error in the $k-$observed ratio.

\begin{figure}
\centering
      \includegraphics[width=0.48\textwidth]{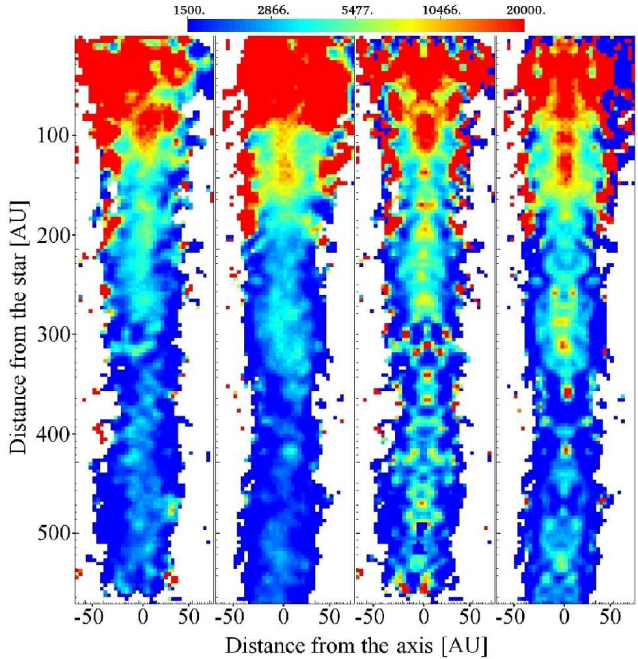}
\caption{Electron density for the first and second epoch for 
         the original (left panels) and reconstructed emission 
         line images (right panels). The x-axis represents the
         direction perpendicular to the jet axis (with a size
         of 60~AU), while the y-axis represents the direction
         parallel to the jet axis (500~AU). The source is 
         located at the top of the Figure.}
\label{fig:ne}
\end{figure}

\begin{figure}
\centering
      \includegraphics[width=0.48\textwidth]{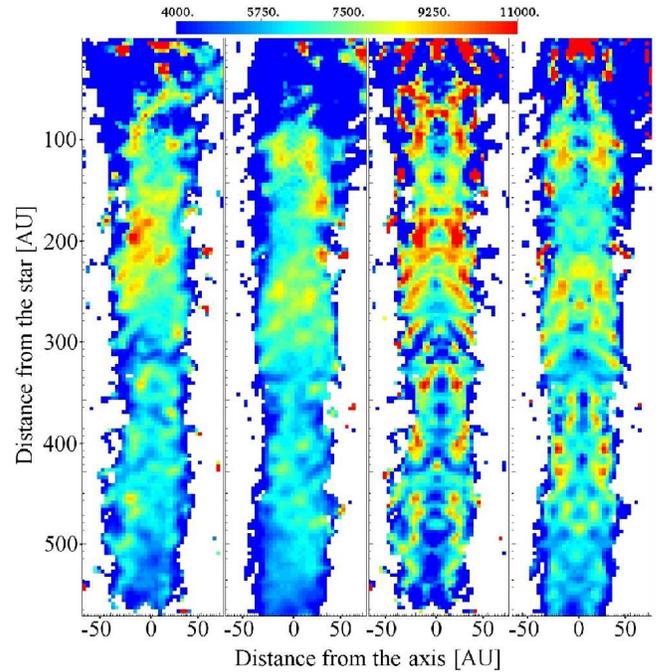}
\caption{The same as in Fig. \ref{fig:ne}, but for the electron
         temperature.
         }
\label{fig:te}
\end{figure}

\begin{figure}
\centering
      \includegraphics[width=0.48\textwidth]{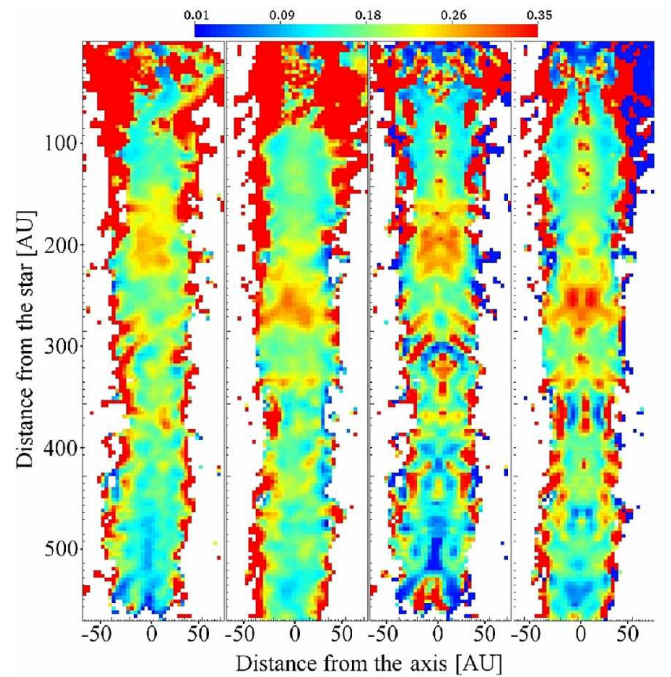}
\caption{The same as in Fig. \ref{fig:ne}, but for the ionization
         fraction.
         }
\label{fig:xe}
\end{figure}

\begin{figure}
\centering
      \includegraphics[width=0.48\textwidth]{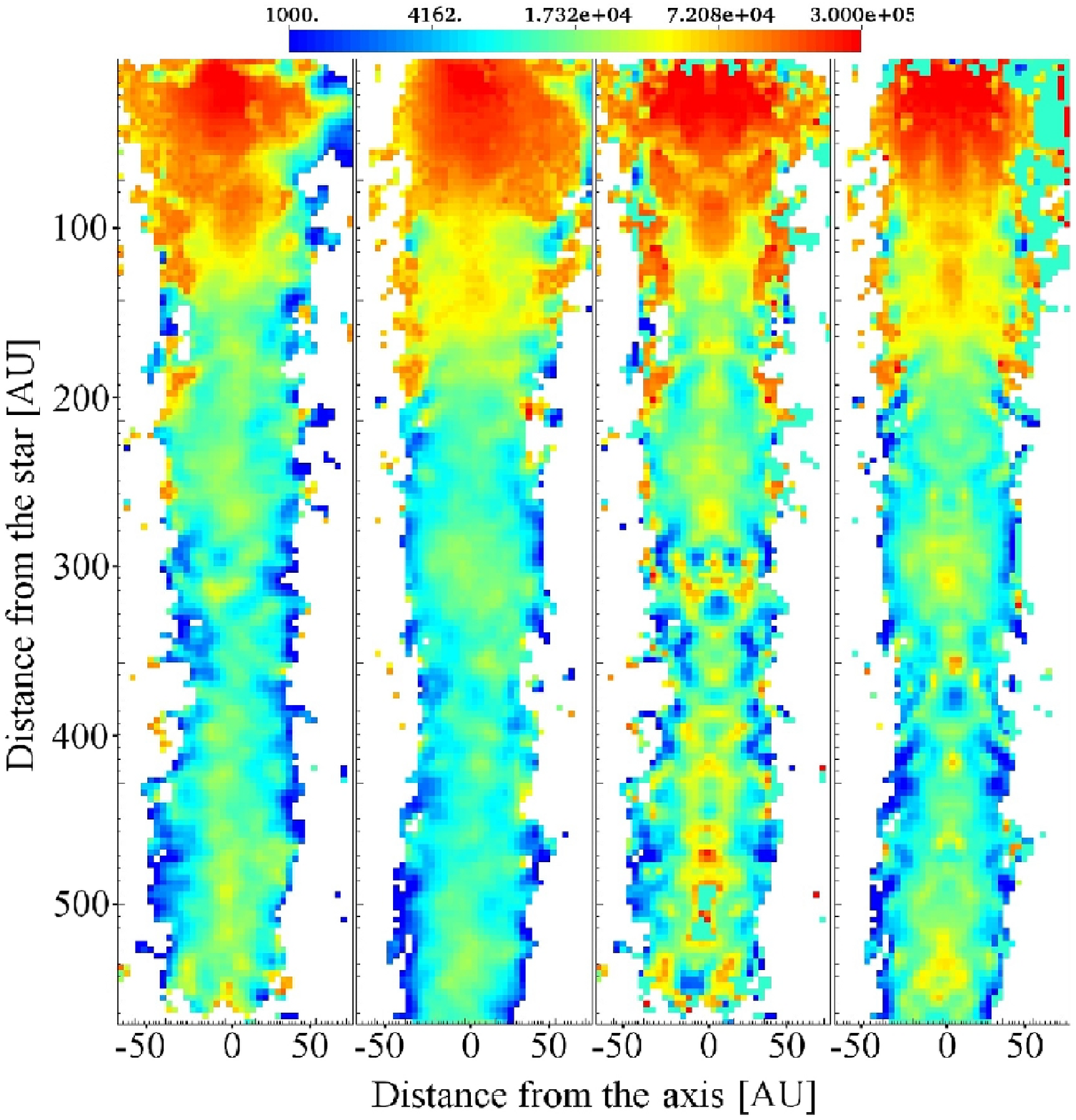}
\caption{The same as in Fig. \ref{fig:ne}, but for the hydrogen
         density ($n_\mathrm{H}$/$n_\mathrm{e}$).
         }
\label{fig:nh}
\end{figure}

The electron density, temperature and ionization fraction
obtained from the observed and inverted ratios are shown in fig. 
\ref{fig:ne}-\ref{fig:xe} for the two epochs, while the derived total
hydrogen density is shown in fig. \ref{fig:nh}. 

In general, the reconstructed $n_\mathrm{e}$, $T$ and $x_\mathrm{H}$
cross sections are much steeper than the observed cross sections
of the same parameters.
The [\ion{S}{2}] ratio is in the high density limit close to the source,
and the electron densities in that region are $\gtrsim 2 \times 10^4$~cm$^{-3}$.
This region can be identified as the saturated region in Fig. \ref{fig:ne}.
While both the observed and the reconstructed electron densities drop 
with the distance from the source, the detailed
behavior is quite different. In fact, the reconstructed electron density
presents a series of stronger on-axis peaks with $n_\mathrm{e} \gtrsim 10^4$~cm$^{-3}$
extending along all the jet.
In those regions, the ionization fraction also increases, while the temperature 
tends to present on-axis valleys. 
Corresponding to lower values of $T_\mathrm{e}$ on the axis, the jet presents regions
of $T_\mathrm{e} \gtrsim 10^4$~K at $\approx 20$~AU from the jet axis.
The inversion process produces total densities ($n_\mathrm{H}=n_\mathrm{e}/x_\mathrm{H}$) 
that are also larger on the jet axis with respect to the 
jet edge, although this difference is lower than the corresponding difference in 
the electron density.
While the $n_\mathrm{e}$ increases by a factor of $\sim 2$ (as in HM07)
from the edge to the axis of the jet for the observed data, 
the reconstructed $n_\mathrm{e}$ increases by a larger factor, of order of
$\sim 5$-$10$ depending on the position along the jet. In general, high
electron density regions are more strongly affected by the reconstruction
process than low density regions.
This is direct consequence of the stronger dependence on density of the
[\ion{S}{2}] ratio close to the high-density limit (see also Fig. 8 of Paper I).

The ionization fraction (both observed and reconstructed) seems to increase 
on the edge of the jet up to values $\gtrsim 0.5$. 
On the edge of the jet, the S/N becomes low and the error in the determination
of the physical parameters becomes dominant, so this result is probably 
affected by numerical uncertainties.
Interestingly, the increase in the ionization fraction on the edges of the jet
is a result expected in shock models, 
as pressure gradients in the post-shock region move the ionized gas laterally in the 
cocoon of the jet, where a region of low density, high ionization fraction is created.
The ionized gas present in that region recombines on a timescale much
larger than on the axis of the jet, where the density is orders of magnitude larger.

\subsection{FWHM}

The degree of collimation of the jet can be estimated by the FWHM.
Typical T-Tauri star jets, including HH30, have an opening angle 
of a few degrees at $\sim 100$~AU from the source, consistent with 
magnetic self-collimation models \citep{2007prpl.conf..231R}.
HM07 measured a deconvolved FWHM of $\approx 10$~AU
close to the HH30 source, with an opening angle of $2.6^\circ$.
HM07 also showed that there is no clear correlation between the jet
collimation/width and the presence of a knot.

\begin{figure*}
\centering
      \includegraphics[width=0.8\textwidth]{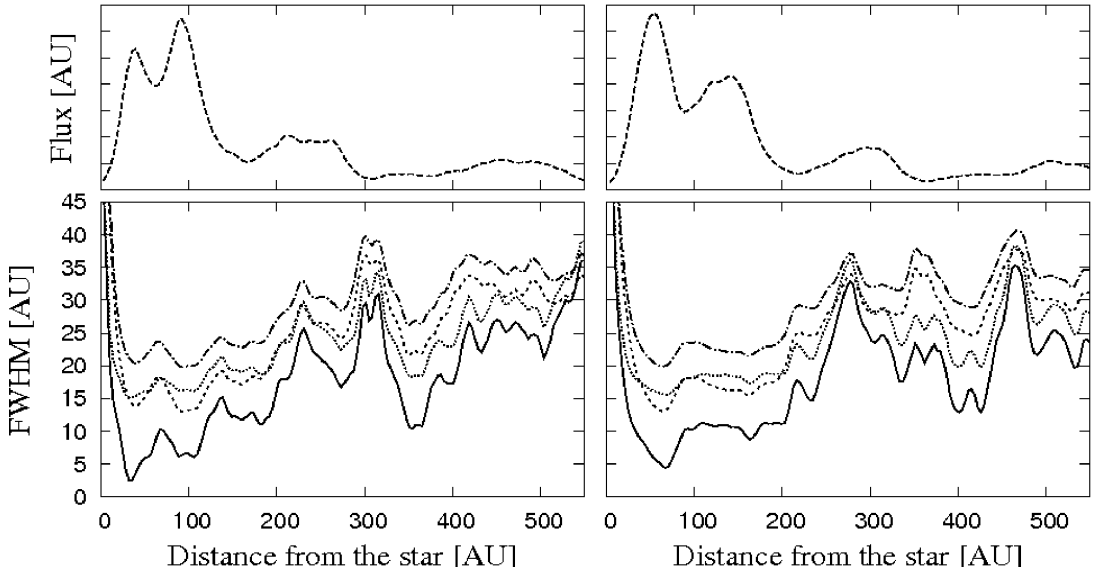}
\caption{\emph{Upper panels}:
Emission line flux (in arbitrary units) as a function
of the distance from the star, for the I (left) and II
epoch (right). The flux is obtained adding 
the [\ion{S}{2}], [\ion{N}{2}] and [\ion{O}{1}].
\emph{Lower panels}: FWHM as a function of distance from the 
star, for the integrated emission line intensities (shown in the upper 
panels of the figure). The different curves correspond to the 
observed (dash-dotted line), reconstructed (dotted), 
deconvolved (dashed) and reconstructed and deconvolved (continuous) data.
The deconvolution has been done assuming a FWHM PSF of 4 pixels.}
\label{fig:fwhm}
\end{figure*}

We have calculated the FWHM before and after the inversion 
(Fig. \ref{fig:fwhm}) for the sum of the [\ion{S}{2}], 
[\ion{O}{1}] and [\ion{N}{2}] emission lines. We have also deconvolved
the FWHM determined from the original and the reconstructed data, assuming 
a FWHM PSF of 4 pixels (HM07). 
The deconvolution of the reconstructed data gives just an approximated 
estimate of the real FWHM, as the deconvolution should be applied
to the data before the reconstruction is made and not after
(see eq. [\ref{eq:abelap1}]).

Fig. \ref{fig:fwhm} shows that results similar to HM07
are obtained for the FWHM calculated from the observed 
and the deconvolved images.
Also, Fig. \ref{fig:fwhm} shows that the inversion and 
the deconvolution produce a comparable effect on the FWHM, 
which has a similar amplitude at small distances from the 
source, going from a width of $\sim 15$~AU at 20 AU to
$30$~AU at a distance of 500 AU from the source. At larger distances 
the FWHM computed from the image inverted by tomography is slightly 
lower than that obtained from the deconvolved image.
The value of the FWHM determined by deconvolving the
reconstructed image is much smaller, going for instance from 5 AU at 30 AU
to 25 AU at 500 AU for the first epoch. The opening angle is
2.4$^\circ$, similar to the one determined by HM07.

The reconstruction process also shows some 
correlation between the knots and the width of the jet.
In fact, the first two knots in the first epoch and the first knot
in the second epoch (fig. \ref{fig:fwhm}) clearly correspond
to a drop in the FWHM, while the rest of the knots do not
present any correlation. That seems to imply that the knot material 
is actually ejected more collimated from the star-disk system than 
the material outside the knots. The reason for the loss of correlation 
seems therefore to be related to the presence of collisions between knots.

\subsection{Mass Loss}

Assuming a uniform velocity cross-section of $200$~km s$^{-1}$, 
and using the density stratification coming from the inversion process, 
we have determined the mass-flux by integrating $d\dot{M} \propto r v n dr$. 
The resulting mass-fluxes computed from the original and the reconstructed 
densities are similar. 
They rapidly drop from $10^{-8}$~M$_\sun$ yr$^{-1}$ close to the source to 
$10^{-7}$~M$_\sun$ yr$^{-1}$ at $200$~AU, that are typical values for
jets from T-Tauri stars (e.g., \citealt{1995ApJ...452..736H}, 
\citealt{2000A&A...356L..41L}).
Interestingly, a very different result is obtained assuming 
a power-law profile for the velocity.
If, for instance, $v \propto r^{-\eta}$, values of $\eta\gtrsim 1$ 
gives mass fluxes $\lesssim 10^{-8}$~M$_\sun$ yr$^{-1}$ close to the source, 
and approximately one order of magnitude lower at at distance of $\gtrsim 200$~AU 
from the source. These values of the mass-flux are much less than those obtained 
by using other methods (e.g., \citealt{1995ApJ...452..736H}, 
\citealt{2000A&A...356L..41L}), implying apparently that the 
velocity cross-section has to be nearly flat (i.e. with a velocity gradient
with $\eta\lesssim 0.5$) across the jet cross-section. Alternatively, these results
can imply that the on-axis total density value is much larger than the one 
derived by the reconstruction process.

\subsection{Errors}
\label{sec:err}

There are three main sources of error that affect the results 
of this work: errors due to the hypothesis of axisymmetery,
errors related to the assumptions of the BE method, and
errors due to the PSF convolution. They are discussed in some detail
in the following.

The first source of errors is related to the hypothesis of axisymmetry.
To understand how asymmetries present in the HH30 jet affect our results, 
we have determined the physical parameters for both sides of the jet 
independently (see \S \ref{sec:lin}). The physical parameters obtained 
differ from those shown by less than $30\%$, without any systematic 
effect present.

The second source of errors is related to the approximations
used in the BE method.
In particular, the main errors in the BE method are related to the uncertainty 
in the elemental abundances.
\citet{2006A&A...456..189P} have shown that the choice of elemental
abundances affects the derivation of the physical parameters
up to 40\% for $x_e$ and 25\% for $T_\mathrm{e}$. On the other side, 
the determination of $n_\mathrm{e}$, coming from the ratio of sulfur
lines, is not affected by the choice of the relative abundances.
Furthermore, the errors in the determination of the electron density
are larger close to the star, where the [\ion{S}{2}] 6716/6731 ratio
is in the high density limit.

The third source of errors is related to the PSF.
In this work we have not deconvolved the data with the PSF, for two reasons.
First, the exact form for the PSF is not available. Second, 
and more important, the mathematical problem becomes 
much more complex if the PSF deconvolution is included (eq. [\ref{eq:abelap1}]).
As we have seen in the determination of the FWHM, the effect
of the PSF is of the same order as the effect produced by the 
inversion process itself close to the source, and is of lower importance
at larger distance from the source. Anyway, the inversion process itself is limited by
the presence of noise in the data. A large level of noise implies 
the need to use a larger regularization parameter. In other words,
the presence of a certain level of noise implies a loss of information,
that cannot be recovered in the inversion process.
However, the PSF deconvolution would produce effects 
in the same direction as the inversion process, resulting in steeper profiles, 
with larger densities and ionization fraction, and lower temperatures on-axis.
The results of this paper remain still valid at least qualitatively, 
and represent a huge improvements with respect to previous works
in which neither the PSF deconvolution nor the line-of-sight integration were
considered in the determination of the physical parameters.


\section{Discussion}

The main results of this paper are that the 
inverted electron density is in general larger, and
that it presents a series of sub-structures not clearly evident 
in the $n_{\mathrm{e}}$ determined directly from the observed data.
The same effect is evident in the total density, although in lower measure, as
the increase in electron density corresponds to a (lower) increase in the 
ionization fraction.

This result has important implications for our understanding of
the origin of HH objects.
The presence of knots in stellar jets seems to be related to the 
generation of internal working surfaces due to supersonic variations 
of the ejection velocity at the base of the jet \citep{1990ApJ...364..601R}.
On the other side, the morphology of the HH30 jet is not easily modeled in terms 
of variable hydrodynamics jet models, because
a periodic velocity ejection from the central source is more likely to produce
knots with ``bow-shock-like'' morphology (e.g., as seen in HH 111, \citealt{1996A&A...311..989R}).
A possible explanation has been proposed by 
\citet{2006A&A...449.1061D}, who showed that axially elongated
structures such as the ones observed in the HH30 jet 
can be produced by pinch instability driven by strong 
toroidal magnetic fields (possibly coexisting with the presence of
internal working surfaces).
The results of this paper suggest a possible alternative explanation.
In fact, the presence of sub-structures in the elongated knots 
indicate that a velocity variation with lower timescales,
maybe of chaotic nature, can be operating, superimposed to the large
scale velocity variation responsible for the creation of knots.
The scenario of chaotic ejection of material forming
the finally visible jet knots has recently been explored in numerical
simulations by \citet{2009ApJ...695..999Y} and
\citet{2010A&A...511A..42B}.

The reconstructed jet presents
regions of lower temperature and higher ionization fraction.
In particular, lower temperature on-axis regions are
qualitatively consistent with shock models, due to 
the stronger cooling that one would expect
in higher density regions.

The same technique could in principle be applied to other existing data, 
although other jets have in general inclination 
angles different from zero and, as a consequence, the interpretation of the
results would be more complex
(e.~g. \citealt{2000ApJ...537L..49B, 2002ApJ...576..222B, 2004ApJ...604..758C, 
2005A&A...432..149W, 2007ApJ...663..350C, 2007AJ....133.1221B}, HM07).

The HH30 data presented in HM07 have a very high S/N level. While the same 
technique can in principle be applied to data with higher uncertainties,
our ability to understand the structure of the jet is 
limited by the noise level of the observations, as the regularization
techniques give smoother solutions for data with lower S/N.

Observations obtained with integral field spectroscopes
(i.e. three-dimensional spectral data with two-dimensional spatial coverage), in
particular, can be used to extract information also on the dependence 
of the velocity on $r$. In this case, in fact, the relation between the observed
intensity and the emission coefficient is given by
\begin{equation}
  I(x,v) = 2\int_x^{\infty} \frac{i(r,v)r dr}{\sqrt{r^2-x^2}},
  \label{eq:ixv}
\end {equation}
with
\begin{equation}
  i(r,v) = {i(r) \over \sqrt{\pi} \sigma_v} e^{-(v-v(r))^2/\sigma_v^2} \,;
  \label{eq:ixv2}
\end {equation}
where $\sigma_v$ is the velocity dispersion.
Given the observed line intensity $I(x,v)$, $i(r,v)$ can be determined 
from equation (\ref{eq:ixv}), $i(r)$ can be determined by 
inverting the Abel equation (related to eq. [\ref{eq:ixv2}] by 
$I(x)=\int I(x,v) dv$) and $v(r)$ can be finally inferred from equation (\ref{eq:ixv2}). 
In practice, $\sigma_v$ is often dominated by the instrumental PSF
(e.g. $10-20$~km s$^{-1}$ vs. a thermal width of $\sim$ a few km s$^{-1}$ for ions as 
\ion{S}{2} or \ion{O}{1}).
On the other side, often $v \gg \sigma_v$ ($\sim$ $100$~km s$^{-1}$ both
across and along the jet) and important information about the 
dynamics of the jet can be determined from $v(r)$, as for instance
the mass-flux.

In cases when information on the width of the jet in two different lines 
(i.e. the [\ion{S}{2}] doublet) is available, equation (\ref{eq:simpl}) can be used 
(see also eq. [18] of Paper I) to have an order of magnitude
estimation of the effect of projection effects.


\section{Conclusions}

In this paper we have for the first time applied the tomographic 
inversion technique to determine, under the hypothesis
of axisymmetry, the three dimensional structure 
of the HH30 jet.
This hypothesis describes better the three dimensional structure 
of the jet than the commonly used assumption of homogeneity
along the line of sight.

The main results are summarized as follows:
the reconstructed electron and total densities show a fragmented structure, 
probably consistent with small timescale ($\sim$ months)
ejection velocity variations.
Corresponding to the peaks in electron and hydrogen density, the ionization fraction
also increases, while the electron temperature presents on-axis valleys.
The width of the reconstructed jet is lower than the corresponding width
inferred directly from the observations. In particular, projection and instrumental
effects have a similar effect on the FWHM. Combining the two effects, the
FWHM becomes $\lesssim 10$~AU close to the source.
Although our determination of the mass-flux is limited by the lack of information
on the velocity radial stratification, we have shown that the effect of the 
density stratification on the mass-flux evaluation is negligible.

The application of the BE technique to the HH30 jet is limited 
to distances from the source $\gtrsim 40$~AU, where the [\ion{S}{2}] 6716/6731 ratio
is not in the high density limit, and is possible to properly determine
the electron density.

Future observations dedicated to the study of the jet close to the source, 
using line ratios with larger critical densities, may allow a direct comparison
between the results of the inversion and predictions of jet ejection models.
In this way, it will be possible to compare for example the density
profiles obtained from the inversion with theoretical predictions. 
For instance, the disk-wind and X-wind predict different power-law dependencies 
of the density as a function of the distance from the jet
axis, while the magnetic tower model predicts a sharp outer jet edge
(e.g. \citealt{1995ApJ...455L.155S,2000ApJ...533..897L}).

We have shown that tomographic reconstruction can be used to reconstruct 
the cross section of the flow, obtaining a description of 
the three dimensional structure of the jet. The use of high
resolution observations will make it possible to apply this 
technique to a larger sample of objects.


\acknowledgments

We thank Pat Hartigan for sharing with us his HST-HH30 data.


\appendix

\section{Implementation of the Tikhonov Regularization}

Here we present the details of our implementation.
For a more general description of the Tikhonov regularization technique
see for instance \citet{tombook,1986ipag.book.....C}.

The observed intensity $I(x_i,y_j)$ is related to the emission coefficients 
$i(r,y)$ by the following expression:
\begin{equation}
  I(x_i,y_j) = 2\int_{x_{i-1/2}}^{x_{i+1/2}} dx
                \int_{y_{j-1/2}}^{y_{j+1/2}} dy
                \int_{-\infty}^{+\infty}
                \int_{-\infty}^{+\infty}
                \psi(x-x^\prime,y-y^\prime) dx^\prime dy^\prime
               \int_{x^\prime-x_0(y)}^R
              \frac{i(r,y) r dr}{\sqrt{r^2-(x^\prime-x_0(y))^2}} \;.
  \label{eq:abelap1}
\end {equation}
The first two integrals of equation (\ref{eq:abelap1}) represent the integration 
on the pixel centered on $(x_i,y_j)$ with dimension 
$(x_{i-1/2},x_{i+1/2}) \times (y_{j-1/2},y_{j+1/2})$.
The third and fourth integrals are the convolution with the seeing and the 
instrumental response $\psi$, and the last term represents the integration 
along the line of sight, being $x_0(y)$ the position of the 
center of the jet (in general a function of the $y$ coordinate).

To simplify the problem, we neglect the effect of the 
instrumental response, assuming $\psi(x-x^\prime,y-y^\prime) \approx \delta(x-x^\prime,y-y^\prime)$,
being $\delta(x)$ the delta function.
Furthermore, we reduce the problem to one dimension assuming $i(r,y) \approx i(r,y_j)$, 
i.e. that the variations
in the emission coefficients along $y$ are much smaller than those
along $r$.

With these approximations, equation (\ref{eq:abelap1}) reduces to
\begin{equation}
  I(x_i,y_j) = 2 \Delta y_j \int_{x_{i-1/2}}^{x_{i+1/2}} dx
               \int_{x-x_0(y_j)}^R
              \frac{i(r,y_j) r dr}{\sqrt{r^2-(x-x_0(y_j))^2}} \;,
  \label{eq:abelap}
\end {equation}
where $\delta y_j = y_{j+1/2}-y_{j-1/2}$. 
We further assume $\Delta y_j = 1$
(the exact value of $\Delta y_j$ is not important, as it cancels out
in the line ratios), and we indicate $I(x_i,y_j)=I_i$ and $i(r,y_j)=i(r)$.

As shown in \S \ref{sec:met}, equation (\ref{eq:abelap}) is inverted 
using the Tikhonov regularization technique,
i.e. solving the following damped least square problem
\begin{equation}	
\min \left(\|A x- d\|_2^2 + \alpha^2 \| L x \|_2^2 \right)
\label{eq:tikA}
\end{equation}
with the condition
\begin{equation}
\|A x- d\| \le \delta \,
\label{eq:alpA}
\end{equation}

In matrix form, equation (\ref{eq:tikA}) can be written as
\begin{equation}
(A^T A+\alpha^2 L)x=A^T d
\label{eq:tra}
\end{equation}
where $A^T$ is the transpose matrix of $A$.
The solution is obtained inverting 
this linear system of equations.

As discussed in \S \ref{sec:met}, the matrix $L$
can be chosen to minimize the norm of the solution, or
its first or second derivative.
In the first case (zero-order) solutions with low magnitude
of $x$ are favored, while in the cases of first- and second- order 
regularization, the magnitude of the gradient and Laplacian 
are penalized, respectively.
Being the solution unknown, the choice of the particular form of 
$L$ is somehow arbitrarily. In this work, we decide to minimize 
the first derivative of the solution as, in our implementation, 
it produces a less noisy solutions with respect to zero- and 
second-order minimization, and it works better in our tests.
The matrix $L$ is therefore given by:
\[ L =  \left| \begin{array}{ccccc}
-1 & 1 & \dots & 0 & 0 \\
0 & -1 & \dots & 0 & 0 \\
\dots & \dots & \dots & \dots & \dots \\
0 &  0 & \dots & -1 & 1 \\
0 & 0 & \dots  & 0 & -1 \end{array} \right|\] 

To complete the description of the algorithm, we need to specify
the form of the matrix $A$.
Let us assume that we have $N$ values of $I_k$, 
defined in the positions $x_k$ (with $0 \le k \le N-1$, where $x_0 = \Delta x_0/2$).
The external radius is defined as $x_{N+1/2}=R$.
Equation (\ref{eq:abelap}) can be therefore written as:
\begin{equation}
  I_k = 2 \int_{x_{k-1/2}}^{x_{k+1/2}} dx 
\left[ \int_x^{x_{k+1/2}} \frac{i_k(r) r dr}{\sqrt{r^2-x^2}}  
+ \sum_{j=k+1}^{N}\int_{x_{j-1/2}}^{x_{j+1/2}} \frac{i_j(r) r dr}{\sqrt{r^2-x^2}} \right]
 \label{eq:exp}
\end {equation}
Taking $i_k(r)=i_{0,k}$ constant inside each shell, this equation leads to:
\begin{align}
  I_k = i_{0,k} 
\left(S_{k+1/2}^{k+1/2}-S_{k-1/2}^{k+1/2}\right)
+ \sum_{j=k+1}^{N} i_{0,j} 
 \left(
 S_{k+1/2}^{j+1/2}-S_{k+1/2}^{j-1/2}-S_{k-1/2}^{j+1/2}+S_{k-1/2}^{j-1/2}
\right)
\label{eq:ik}
\end {align}
where 
$S^n_m=x_m\sqrt{x_n^2-x_m^2} + x_n^2\arcsin{x_m/x_n}$, and 
the following relation has been used:
$ 2 \int \sqrt{x_n^2-x^2} dx = x \sqrt{x_n^2-x^2}+x_n^2 \arcsin{x/x_n} = S^n$.
The matrix $A$ can be easily deduced from equation (\ref{eq:ik}).

\begin{figure}
\centering
      \includegraphics[width=0.7\textwidth]{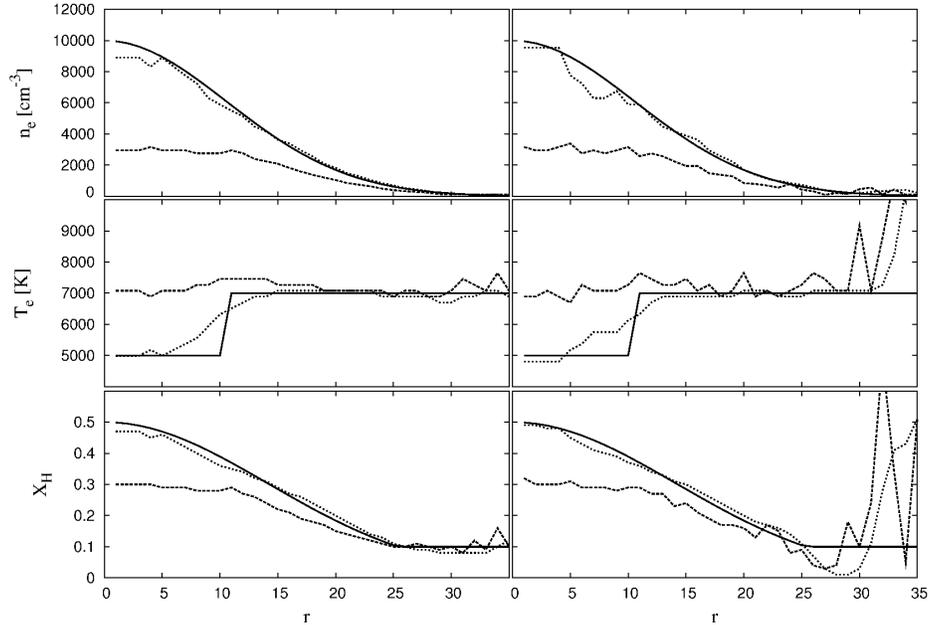}
\caption{Test of the Tikhonov reconstruction technique 
         for a Poisson noise with a variance of 5\% (left column) and
         10\% (right column) at $r=5$.
         The different panels correspond to $n_\mathrm{e}$ (upper panels), 
         $T_\mathrm{e}$ (center), $x_\mathrm{H}$ (lower) respectively.
         In each panel, the different curves represent the original profile 
         (continuous line), that obtained from the integrated 
         intensities (dashed line), and the reconstructed profile (dotted line).
         The reconstructed profiles is less steep than the original. The step
         in temperature is not well reconstructed by the Tikhonov technique.
         The edge of the jet is dominated by noise. The noise level of the 
         physical parameters obtained from the integrated and reconstructed
         profiles are similar.}
\label{fig:test}
\end{figure}

We have run a series of tests to verify the accuracy of the algorithms developed in this paper.
Fig. \ref{fig:test} shows one of these tests.
Assuming a certain radial dependence for electron density, temperature and
ionization fraction, we have computed synthetic emission line intensities.
We have then integrated the volumetric emission line along the line of sight
using equation (\ref{eq:abelap}), and then added a Poisson noise to the data.
Finally, we have reconstructed the electron density, temperature and ionization fraction
using the tomographic technique and the BE method (\S \ref{sec:par}).
As shown in Fig. \ref{fig:test}, the resulting physical parameters agree with the original
one within $10\%$. For comparison, we show in Fig. \ref{fig:test} also the physical
parameter cross-sections obtained assuming the medium as homogeneous along
the line of sight (i.e., those obtained from the ``observed'' emission lines).
The tests show that also with a $~10\%$ noise level in the data our implementation
of the tomographic inversion gives acceptable results.




\end{document}